# Unveiling cells' local environment during cryopreservation by correlative *in situ* spatial and thermal analyses

Kankan Qin[a], Corentin Eschenbrenner[a], Felix Ginot[b], Dmytro Dedovets[b], Thibaud Coradin[a], Sylvain Deville[b, c], Francisco M. Fernandes[a]*

[a] *Sorbonne Université, UMR 7574, Laboratoire de Chimie de la Matière Condensée de Paris, F-75005, Paris, France*

[b] *Laboratoire de Synthèse et Fonctionnalisation des Céramiques, UMR 3080 CNRS/Saint-Gobain CREE, Saint-Gobain Research Provence, Cavaillon, France*

[c] *now with: Université de Lyon, Université Claude Bernard Lyon 1, CNRS, Institut Lumière Matière, 69622 Villeurbanne, France*

*Abstract*

Cryopreservation is the only fully established procedure to extend the lifespan of living cells and tissues, a key to activities spanning from fundamental biology to clinical practice. Despite its prevalence and impact, central aspects of cryopreservation, such as the cell's physico-chemical environment during freezing, remain elusive. Here we address that question by coupling *in situ* microscopic directional freezing to visualize cells and their surroundings during freezing with the freezing medium phase diagram. We extract the freezing medium spatial distribution in cryopreservation, providing a tool to describe the cell vicinity at any point during freezing. We show that two major events define the cells' local environment over time: the interaction with the moving ice front and with the vitreous moving front – a term we introduce here. Our correlative strategy may be applied to cells relevant in clinical research and practice, and help designing new cryoprotective media based on local physico-chemical cues.

*Introduction*

For decades, cryopreservation has been the pivotal strategy in preserving functional living matter for extended periods of time. The capacity to overcome the lifespan of individual cells, cell assemblies or tissues has enabled much of the recent advances in fundamental cell biology, cell therapies and tissue engineering.[1] This ability has been key for several clinical breakthroughs in reproductive biology[2], and more generally all cases where the use of immortalized cell lines is not suitable. Cryopreservation is generally attained via two – seemingly opposite – approaches whose common purpose is to prevent the deleterious effects, direct or not, of water solidification on cells: *vitrification* and *slow freezing* methods.[1] *Vitrification* consists in a liquid-solid transition of water – intra- and extracellular – to an amorphous (vitreous) state by rapidly cooling (up to $10^6$ °C.min$^{-1}$)[3] below the glass transition temperature[4], preventing ice crystallization and thus minimizing cryoinjury. Conversely, *slow freezing* methods, that involve programmed temperature decrease, are nowadays widely applied for cryopreservation of cells and tissues[5,6]. Beyond their technical differences, both *vitrification* and *slow freezing* strategies rely on an entangled set of thermodynamic and kinetic parameters that blur the analysis of the freezing phenomena. In addition, their respective success is determined by the viability and/or vitality of cells and tissues that are frozen in presence of permeating cryoprotectant agents (pCPAs) such as DMSO which may impose cytotoxic response by cells.[7–10] The challenges in establishing causal effects between the physico-chemical freezing parameters and the biological responses have diverted the main research efforts towards a different direction, *i.e.* formulating new freezing media that maximizes cell viability to the stresses imposed by freezing and thawing.

Strategies to inhibit or minimize ice crystals formation while reducing the harmful effects of most pCPAs (*e.g.* DMSO, glycerol) were proposed by lowering the later concentrations in presence of non-permeating cryoprotectant agents (npCPAs).[11] Also sucrose and trehalose were found to accelerate dehydration of cells by altering intra- and extracellular osmotic pressures, which to some extent reduces the possibility of intracellular ice growth.[12,13] Recently, antifreeze proteins (AFPs) that present ice recrystallization inhibition (IRI) properties have been applied as freezing media[14], but immunological and toxicological issues are still concerns for their clinical use[15,16]. To circumvent these issues AFP mimics such as Poly(vinyl alcohol) (PVA) have been used for animal cell[17] and bacteria[15] cryopreservation as biocompatible alternatives endued with IRI properties. Complex carbohydrates such as hydroxyethyl starch (HES)[18,19], alginate[20–22] or synthetic polyampholytes[23] have been adopted in the field of cell cryopreservation either with other pCPAs or alone. More radical solutions have also been proposed such as the use of metal-organic frameworks,[24] armors around cells to prevent damage by ice crystals or hydroxyapatite nanoparticles,[13] as vehicles to promote the internalization of CPAs. In most of the preceding cases, may they be classical solutions such as DMSO or more sophisticated solutions such as AFPs, one crucial question remains open. What is the local composition of the vicinity of cells during freezing?

Here we address that question by using a very simple freezing medium, sodium alginate solution in water, to focus on the cellular environment during the freezing process. We adapted the directional freezing method developed by Dedovets *et al.*[25] to investigate the interaction between suspended *S. cerevisiae* cells and a moving freezing front in presence of a polysaccharide salt solution. These observations are combined with the analysis of the water-sodium alginate phase diagram obtained by Differential Scanning Calorimetry (DSC) to draw a more complete picture of the freezing environment experienced by *S. cerevisiae* cells throughout the freezing process. Directional freezing has been previously applied to observe red blood cells during freezing[26] and to preserve adherent cells[27] as well as ovarian[28], liver[29] and heart[30] tissue in complex pCPA-containing media. In this work directional freezing provides a controlled freezing environment with clearly defined thermal boundary conditions, steady freezing rate, and the ability to investigate *in situ* the interaction between cells and the freezing front. These conditions enable to bridge the kinetic and thermodynamic events occurring during freezing, to reach a more complete view of *S. cerevisiae* environment prior, during and after interaction with the freezing front. Moreover, we demonstrate that despite the slow freezing approach and the absence of pCPAs, the effective segregation between pure ice crystals, and an increasingly concentrated phase, rich in cells and alginate, leads to a cellular local environment characterized by a vitreous behavior whose predominance correlates positively with yeast viability.

*Thermodynamics of freezing media*

Almost all compounds soluble in water are practically insoluble in hexagonal ice ($I_h$). This implies that most solutes are segregated from ice crystals during water freezing.[31,32] Under specific conditions, also particles[33] and even cells[20,26,34] in suspension segregate towards the interstitial space defined by the newly formed ice crystals. In the following we estimate cells to stay in the interstitial space, and verify this assumption experimentally. The consequence of such segregation step is central in cryobiology since it determines the local composition of the cellular environment during freezing, an information that was experimentally unavailable and conceptually overlooked.

Figure 1.a illustrates the path of an individual cell (in cyan) at two key moments that are defined by the solute concentration changes. To characterize the increasingly concentrated media generated during ice growth a binary phase diagram of sodium alginate in water was established using DSC (Figure 1.b). The freezing temperature $T_f$ was determined at the onset of the exothermic crystallization peak for polymer volume fractions, $\phi_p$ ranging from 0.025 to 0.65 in consecutive freezing and heating cycle at 10 °C.min$^{-1}$ (Figure S1). According to Flory's theory, the freezing point of a polymer solution with high polymerization degree ($N >>> 1$) can be described by equation 1,[35]

$$T_f = \frac{1+\frac{w}{\Delta H}\phi_p^2}{\frac{1}{T^0}-\frac{R}{\Delta H}(\ln(1-\phi_p)+\phi_p)} \qquad (eq.1)$$

where $w$ is the regular solution interaction parameter in $J.mol^{-1}$, $\Delta H$ is the water melting enthalpy ($\Delta H = 6007\ J.mol^{-1}$), $\phi_p$ is the polymer volume fraction and $R$ is the Boltzman constant in $J.K^{-1}$. For $\phi > 0.54$ no exothermic peak corresponding to water freezing was detected at -10 °C.min$^{-1}$, suggesting that for high polymer content compositions no ice crystallization occurred within the detection limit of the DSC. The lack of detectable freezing events is in good agreement with the glass transition temperatures calculated for the ice-alginate system defined by Gordon-Taylor equation[35] (eq. 2),

$$T_g = \frac{T_{gw}\phi_w+kT_{gp}\phi_p}{\phi_w+k\phi_p} \qquad (eq.\ 2)$$

where $T_g$ corresponds to the glass transition temperature for a given composition, $T_{gw} = -137.15\ °C$ is the pure ice glass transition temperature, $T_{gp} = 120°C$ is the polymer glass transition temperature and $k$ is an empirical parameter fixed to 0.5 as discussed elsewhere for other polysaccharide systems[35]. The full phase diagram of the alginate/water binary system is depicted in Figure 1.b. As reported for similar polysaccharide systems[35], upon cooling at 10 °C.min⁻¹, sodium alginate solutions (L) below $\phi < 0.54$ tend to segregate into a pure ice phase (I) along with an increasingly concentrated solution phase (L). If the mixture is cooled below the temperature defined by the intersection between $T_f$ and $T_g$ then a vitreous (V) or glassy phase is formed whose composition is defined by the same intersection between $T_f$ and $T_g$.

From another set of DSC experiments we quantified the ice volume fraction $\phi_{ice}$ from the integral of the ice melting enthalpy in alginate-water binary mixtures. From the values of $\phi_{ice}$ obtained, we determined the volume fraction of non-frozen water associated with the polymer for each alginate concentration. Figure 1.c depicts the evolution of the non-frozen water volume fraction ($local\ \phi_{water}$) and alginate volume fraction ($local\ \phi_{alginate}$) in between ice crystals after freezing different alginate-water binary mixtures from room temperature to $-80\ °C$ at $10\ °C.min^{-1}$. The local composition in between ice crystals was found to be independent of the initial alginate-water composition up to $\phi = 0.6$ (dark grey and magenta data points). Above this alginate volume fraction no ice melting endotherm was detected upon heating the samples from -80 °C to 40 °C, suggesting that no ice crystallization took place during freezing. Under these circumstances all water was considered as non-freezing water (a thorough discussion of the nature of freezable and non-freezable water in presence of polysaccharides is available elsewhere and is beyond the scope of this work[35]). As a consequence, no phase segregation occurred and the local composition of the polymer-rich phase after freezing was similar to the initial mixtures (lighter grey and magenta data points). The intersection between the liquidus curve and the glass transition curves (Figure 1.b) thus defines the local composition of the segregated phase for $\phi < 0.6$. This critical polymer volume fraction will be referred to hereafter as $\phi_{crit}$. For all practical uses in cell cryopreservation, polysaccharide concentrations is lower than $\phi_{crit}$. The consequences from a cryopreservation standpoint are important since these data suggests that regardless of the initial concentration of polymer (provided that $\phi < \phi_{crit}$), the composition of the local environment surrounding cells is independent of the initial polymer concentration and is defined solely by the intersection between $T_f$ and $T_g$.

Since the initial composition of alginate solutions we used in the cell cryopreservation experiments was limited to $\phi = 0.025$ (vertical dotted line in Figure 1.b), and because the thermal gradient in the directional freezing setup is linear, it is possible to transpose the data from the phase diagram to establish the composition of the cell surroundings from room temperature to -80 °C during freezing. Such an analogy enables to describe both the temperature and composition of the ice/alginate solution/vitreous phase throughout the full freezing process. We thus define two moments that separate different polymer concentration around individual cells during freezing. At $t_1$ (Figure 1.a) the cell surroundings evolved from a fixed polysaccharide concentration defined by the initial water-alginate mixture to experience an increase in local polymer concentration until this concentration reached $\phi_{crit}$ at $t_2$ (Figure 1.a). At $t_2$ alginate concentration in the cells' surrounding became constant and equivalent to the vitreous mixture of alginate and water defined by $\phi_{crit}$.

*Local concentration of alginate between ice crystals*

It is commonly admitted that each cell type displays optimum viability at a specific freezing rate. The bell-shaped curves that describe cell viability according to freezing rate, have supported this empirical statement and have provided much needed guidance in the definition of the best cryopreservation protocols for different cell types.[36,37] The description of the viability according to the freezing rate is, however, provides only limited insight into the different phenomena occurring during cryopreservation. As shown above, the composition of the interstitial zones, and thus the cellular environment during freezing, depends on the intersection point between $T_f$ and $T_g$. The liquidus curve describes a purely thermodynamic transition whose characteristic temperature for a given composition is not expected to change with the freezing rate. The glass transition is, on the contrary, strongly dependent on the heating and cooling rates[38]. Faster cooling rates are expected to move $T_g$ to higher temperatures, which changes the intersection point between $T_f$ and $T_g$ and thus impacts on $\phi_{crit}$, the composition of the vitreous phase. In the following experiments we have controlled the ice front velocity as a strategy to modulate the freezing rate in the directional freezing setups. To ascertain the impact of ice front velocity on the phase segregation between ice and alginate/water glass during freezing, an alginate solution marked with rhodamine B was frozen under the confocal microscope setup depicted in Figure 2.a at 10, 20, 30 and 50 µm.s$^{-1}$. Two dimensional image sequences of the freezing phenomena were acquired and stabilized at the moving front. Figure 2.b represents a section of a still image used to calculate the volume fraction of ice according to the distance to the ice front for the different ice front velocities. At least ten consecutive images were treated (Figure S2) for each ice front velocity. From the integration of the ice zones (in black) the ice volume fraction as a function of the distance to the ice front was determined. Figure 2.c describes the progression of the ice volume fraction, $\phi_{ice}$, along the $x$ direction according to the ice front velocity. From the moving ice front point towards the cold element, $\phi_{ice}$ grows rapidly to attain a relatively stable regime at around 400 µm from the ice front. The ice volume fraction at the plateau zone is independent of the ice front velocity for 10 and 20 µm.s$^{-1}$. As the ice front velocity increases from 20 to 50 µm.s$^{-1}$, $\phi_{ice}$ at the plateau decreases from 0.91 to 0.79, respectively. While this difference may seem of little relevance to understand the local environment of cells during freezing, the conversion to local alginate concentration in the interstitial zones is revelatory. In figure 2.d, for highest ice front velocity (50 µm.s$^{-1}$), the local alginate concentration evolves from 4 wt.% to attain a stable regime at around 20 wt.%. For the lower ice front velocities (10 and 20 µm.s$^{-1}$) a mostly linear trend indicates that the local concentration of alginate evolves from 4 wt.% to 40 wt.% within the first 600 µm. The intermediate ice front velocity, 30 µm.s$^{-1}$ yields a mostly linear alginate concentration variation with the distance to the ice front, reaching 26 wt.% at 600 µm. These results clearly indicate that the local environment surrounding cells during directional freezing is largely controlled by the ice front velocity. The difference in alginate concentration profile evidenced in Figure 2.d can be further explained by the fact that the lowest temperature available in the confocal setup is -8 °C. At high ice front velocities the $T_g$ is higher and the intersection between the liquidus line and $T_g$ is readily accessible at moderate temperatures such as -8 °C. This leads to reaching a plateau in composition under the confocal microscope as observed for 50 µm.s⁻¹. At lower velocities the intersection between the liquidus line and $T_g$ can only be accessed at lower temperatures, leading the experiment not to attain the steady state regime. While this limitation is present and relevant in the confocal setup it was fully eliminated in the cell freezing setup used later in this work where the cold plate can go as low as -100 °C.

Another aspect that was accessible from the previous data relates to how the variation of alginate concentration occurs with time, for different ice front velocities, from $t_1$ onwards. Figure 2.e describes the local alginate concentration increase within the accessible field view under the confocal microscope setup, from the ice front to 600 µm. For low velocities, the concentration increase was progressive but reached high alginate concentration (up to 40 wt.%). At higher ice front velocities the alginate concentration increase was sharper but reached a far more limited local concentration (20 wt.%). Figure 2.f depicts the initial concentration increase immediately after $t_1$. The slopes of the linear fits show a marked difference in the initial concentration variation ranging from 1.3 to 7.1 wt%.s⁻¹ for 10 and 50 µm.s⁻¹, respectively.

It is thus reasonable to hypothesize that, in directional freezing, the ice front velocity can play two major roles impacting cell behavior. The first is related with the final composition of the vitreous alginate-water phase surrounding the cells that varies from 20 to 40 wt% by solely modifying the ice front velocity from 50 to 10 µm.s⁻¹, respectively. This may account for important variations in cell viability since the hydration of the storage medium will, in the present water-polysaccharide system in the absence of pCPAs, define the osmotic pressure difference between the intra- and extracellular media of the cells after $t_2$. The second role of the ice front velocity is related with the rate at which the polymer concentration changes from $t_1$ onwards. Assuming that suspended cells will, prior to interaction with the freezing front, be at osmotic equilibrium with the alginate solution ($\phi = 0.025, 4 \ wt\%$), the ice front velocity determines the rate at which cells will be drawn to an osmotic out-of-equilibrium situation.

The local environment of cells during freezing is determined by a diverse range of conditions that extend beyond the local concentration of polymer, as discussed so far. Another particularly important aspect is the fate of cells during their interaction with the freezing front. As reported previously[20], yeast cells are observed covered in a polysaccharide envelope, regardless of the ice front velocity used in the directional freezing (Figure S3). However, to establish a quantitative relationship between cell viability and ice front velocity we should ensure that the ratio of cells that end up within the interstitial space defined by ice crystals (*i.e.* enveloped in alginate) over those being engulfed directly by ice is independent of the ice front velocity.

*Cell viability*

To determine the physical environment of suspended *S. cerevisiae* cells during freezing the home-built directional freezing setup[20] (Figure 3.a) was used to freeze yeast cell suspensions at different ice front velocities. The three dimensional image of the freezing front zone, depicted in Figure 3.b illustrates how ice crystals (unmarked, white) segregate both alginate solution (marked with Rhodamine B (Rh-Alg), coded in magenta) and *S. cerevisiae* cells (marked with FUN1 dye, coded in cyan) to form domains of 4-6 µm in the y direction (after the ice crystal is fully formed, bottom of the image) separated by ice crystals ranging between 20 and 50 µm in the y direction (see Figure S2). From the 3D rendered image, yeast cells appear covered by a magenta layer (rhodamine B in alginate), indicating the absence of direct contact between the cell wall with ice crystals. This observation was further confirmed by a sequential imaging of cells exposed to an ice front moving at 10 µm.s$^{-1}$. Figure 3.c depicts nine sequential fluorescence 2D images of the freezing front before, during and after the interaction of two suspended cells with the moving freezing front. Two different outcomes were identified: encapsulation in the interstitial space defined by ice crystals (white arrows) and engulfment of the cells in the ice crystals (yellow arrows). The images in Figure 3.c were chosen due to the co-occurrence of both cases and are not representative of the whole sample. Statistical analysis of the encapsulation efficiency of yeast cells (n > 200) by the moving ice front (Figure 3.d) performed on image sequences obtained at 10 and 50 µm.s$^{-1}$ yielded more than 85% of cells were encapsulated in the alginate-rich zones, regardless of the ice front velocity.

As shown above, the freezing conditions determine the final polymer concentration and the rate at which this concentration is attained based solely on the velocity of the freezing front. To ascertain if ice front velocity did translate into different cell viability when applied to yeast cell suspensions a dedicated directional freezing setup was built. This setup allowed to control the freezing step between two well-defined thermal boundaries closer to the temperatures relevant in cryopreservation (Figure 4.a). *S. cerevisiae* cells suspended in alginate solution were placed between glass coverslips separated by a 500 µm spacer and moved from the hot block ($T_H = 10\ °C$) towards the cold block ($T_C = -100\ °C$) at different linear velocities. To ascertain the impact of the ice front velocity on the viability of *S. cerevisiae* cells during freezing, three ice front velocities, 10, 50 and 100 µm.s$^{-1}$, were used. Experimental data describing the cooling rate according to the ice front velocity is available in the supporting information (Figure S4). The growth curves of yeast cells were obtained in liquid growth conditions after directional freezing followed by controlled quick thawing to 37°C (Figure 4.b). The viability results support previous findings on the inverse dependency of cell viability with freezing rate.[36,37,39] When frozen at 10 µm.s$^{-1}$ (5 °C.min$^{-1}$) *S. cerevisiae* cell suspensions (1.5·10$^7$ cells.mL$^{-1}$) display significantly faster growth kinetics than at higher ice front velocities (and freezing rates). Though significantly different from the control up to 32h, the characteristic growth curve of cell suspensions frozen at 10 µm.s$^{-1}$ follow the control growth curve closely. This behavior is also observed for different cell densities ranging from 1.5·10$^6$ to 1.5·10$^8$ cells.mL$^{-1}$ (Figure S5). As the ice front velocities increased from 10 to 50 and 100 µm.s$^{-1}$ the cells' growth kinetics decreased in a monotonic manner indicating the inverse dependence between cell viability and the physico-chemical conditions determined by the ice front progression. These results were further confirmed by plate counting after a freeze-thawing process (Fig. 4.c). The colony forming units determined by plate counting also decreased monotonically with ice front velocity.

*Discussion*

Establishing a strict causal relationship between yeast viability and the composition of the cell surroundings determined by the ice front velocity is not possible since a wide range of other parameters are simultaneously at play. The pressure formed in between ice crystals during ice templating and its dependence with ice front velocity cannot be discarded. Recent results suggest that the pressure generated during ice templating of a lamellar glycolipid can attain values in the kbar range.[40] Although only two ice front velocities were tested (9 and 15 $\mu$m.s$^{-1}$), the pressure values do stand in the critical range for eukaryotic cells (between 1 and 3 kbar).[41] It is likely that a wider ice front velocity range may translate into significantly different interstitial pressures with repercussions in cell viability. Also the morphology of the ice crystals formed during freezing are markedly different depending on the ice front velocity. The minimum Feret diameter of the macroporous alginate structures observed under SEM after freeze drying, displayed in the supplementary information (Figure S6) and the periodic distances obtained under the confocal microscope (Figure S7) are relevant measures to describe the wall-to-wall distance of the lamellar pores generated by the growth of ice crystals. The inverse scaling between ice crystals size and ice front velocity observed here is a common behavior described for ice-templating of a variety of systems ranging from biopolymers[20,42] to ceramic slurries[43,44]. While the morphological feature may seem of little relevance to cell viability, the size and distribution of ice crystals within the sample may have significant impact during thawing of the samples, notably on their likelihood to endure recrystallization, a key aspect in cell injury.

Nevertheless, we have here minimized the unknown variables usually associated with cryopreservation by selecting a freezing setup with clearly defined boundary conditions that can be reproduced under a confocal microscope to draw a clearer picture of the cellular environment during freezing. We can therefore unveil, in a detailed manner, how the medium surrounding each cell evolves during freezing and how the ice front velocity determines the composition of the cell environment. Based on the water-alginate phase diagram we introduce the notion of *vitreous moving front* as a key element to understand how the medium concentration evolves over time during freezing. Of especial importance is the impact of the ice front velocity on the evolution of the concentration of the polymer solute surrounding cells. Increasing the ice front velocity led to a lower concentration of alginate within the interstitial space where cells are confined. This evolution could lead to a lower osmotic pressure difference between the extracellular and the intracellular spaces and thus to less osmotic stress. In this regard, fast ice front velocity would seem a favorable condition to maximize cell viability. However, an attentive look at the initial solute concentration rate of change between $t_1$ and $t_2$ shows very large variation, up to 7.1 wt%.s$^{-1}$, at high velocity, whereas this value was as low as 1.7 wt%.s$^{-1}$ for slower ones. We hypothesize that this slow rate of solute concentration change is a central parameter behind the increased viability of cells frozen at 10 µm.s$^{-1}$. Higher rates may not allow to reach an equilibrium state between the intra- and extracellular osmotic forces, leading to an excess of cytoplasmic water and therefore inefficient inhibition of intracellular ice growth by solute/crowding effects.

Coupling microscopic observation during directional freezing of cells with the thermal analysis of the freezing medium enabled to characterize the cell local environment during freezing. To the best of our knowledge, describing the composition of the freezing media surrounding cells during freezing has never been accomplished. Here we experimentally quantify the evolution of the concentration of a polysaccharide-based medium during freezing and we establish that its final concentration is independent from its initial concentration in solution below a critical value. We describe the vitreous nature of the polysaccharide surrounding cells as it accumulates in the interstitial space defined by ice crystals. Most transformations in concentration occur between the time points that mediate between the passage of the moving ice front and the moving vitreous front, a concept we introduce for the first time. These observations strengthen the role of controlled ice front velocity in cryopreservation, a factor often overlooked in cooling rate control experiments. Due to the eminently kinetic nature of the vitreous transition, we prove that the different composition of the cell surroundings depends on the ice front velocity used during freezing.

These results provide a more detailed insight into the osmotic stress endured by cells during cryopreservation due to the local increase in medium concentration, an information that has remained elusive up to now. Furthermore, the correlative approach developed here should allow to achieve a deeper understanding of the freezing mechanisms of biological materials and provide quantitative data to rationalize cell survival during cryopreservation, an essential step to achieve better freezing protocols in absence of toxic cryoprotectants.

*Materials and Methods*

*DSC* – Differential Scanning Calorimetry was performed to study the thermal behavior of alginate/water mixtures. In a typical experiment, 20 mg of different composition varying from 4 wt.% to 80 wt.% were sealed in aluminum crucibles and left to homogenize for one week at ambient temperature. To analyse the fraction of frozen water, samples were cooled to -80 °C at –10 °C.min⁻¹ followed by heating back to room temperature. The ice melt endotherm was analyzed to understand the local composition of composition of the final system. The comparison between ice melting endotherm integration and the amount of introduced water allowed the quantification of the amount of non-freezing water associated with the polymer fraction according with the initial alginate concentration. To describe the cryoscopic depression, samples were cyclically cooled to -80 °C and heated back the temperature corresponding to the middle of the melt endotherm. Upon each cycle the return temperature between heating and cooling ramps was shifted of 0.5 °C towards lower temperatures to overcome supercooling. All DSC analysis were performed in a TA Instruments Q20 DSC instrument coupled with a cooling flange. The conversion between mass and volume fraction is according to the equation (S.eq.1) in the supplementary information.

*Directional freezing under confocal microscopy* – The confocal microscopy system designed to inspect the ice front during directional freezing[25] was composed of a temperature controller connected to two independent Peltier elements (ET-127-10-13, Adaptive, purchased from RS Components, France) coupled with a controlled XY stage. The distance between the thermal elements was 2 mm. The freezing process was observed under a Confocal Laser Scanning Platform Leica TCS SP8 (Leica Microsystems SAS, Germany). For the analysis of the polymer fraction, rhodamine B (0.1 mM) was dissolved in 4 wt% alginate solution. The obtained fluorescent solution was injected inside Hele-Shaw cells. The filled Hele-Shaw cell were moved from the hot to the cold Peltier elements (temperature ranged from 12 to -8 °C) by a stepper motor at 10, 20, 30 and 50 μm.s⁻¹ linear velocities,. The ice crystal and corresponding segregated polymer volume fractions were analyzed with FIJI software[45]. For cell imaging during freezing, *S. cerevisiae* cells (around 1.5·10⁷ cells.mL⁻¹) stained with LIVE/DEAD™ Yeast Viability Kit (L7009, ThermoFisher Scientific) were mixed with alginate solution in presence of Rhodamine B (0.1 mM) for 30 minutes. The samples were then imaged during freezing as they crossed the thermal gradient at 10 and 50 $\mu$m.s$^{-1}$. Image analysis was performed using Fiji software[45]. The encapsulation efficiency of yeast cells (stained by SYTO 9 and propidium iodide, both obtained from Thermo Fisher Scientific) was obtained from counting cells trapped inside of alginate walls or ice column (n > 200).

*Cell culture* – 10 mg of dry *Saccharomyces cerevisiae* cells (obtained from SIGMA-ALDRICH) were suspended in 1 ml pH 7.2 PBS solution. Subsequently 10 $\mu$l of cell suspension were pipetted into 10 ml Yeast Extract Peptone Dextrose (YPD) culture medium. Cell suspensions were then placed in an incubator at 30 °C, 150 rpm for 30 hours. Cells were harvested after centrifugation at 6000 rpm for 10 minutes and resuspended in 0.9 wt.% NaCl solution for the following experiments.

*Directional freezing of yeast cells* – The setup was composed of two temperature-controlled aluminum plates (12×10×1.5 cm) used as cold ($T_C = -100 \pm 5$ °C) and hot element ($T_H = 10 \pm 2$ °C) and a motor to slide the samples between both aluminum plates. The cold plate was refrigerated via a copper element plunged into liquid Nitrogen. The hot plate was controlled by a circulating flow of water from a nearby temperature controlled bath. The distance between the two plates was set to 2 mm. Mixture of 4 wt.% sodium alginate aqueous solution and yeast cells (final concentration, 1.5·10$^7$ cells.ml$^{-1}$) was stabilized for 30 minutes at room temperature and injected into a sample holder (a glass coverslip sealed with a 500 $\mu$m culture chamber from Thermo Fisher Scientific). Prior to freezing, a stabilization period (30 mins. at $T_H$) was systematically applied for the samples to attain thermal equilibrium, upon which the samples moved at 10, 50 or 100 $\mu$m.s$^{-1}$ from the hot to the cold element. Once the sample was fully frozen and reached the edge of the cold element it was stored at -80 °C for 12 h. The frozen samples were fully thawed in a water bath at 37 °C for 15 seconds. To ascertain cell regrowth after freezing and thawing, 120 $\mu$l of each thawed solution was dispersed into 10 ml fresh YPD broth medium and incubated at 30 °C, 150 rpm. Aliquots were drawn at 8, 12, 24, 28, 32, 48, 56 h and optical density at 600 nm was tracked using a UV spectrometer (UVIKONXL SECOMAM). Three replicates were performed for each ice front velocity. Directional freezing was performed at different cell densities, ranging from 1.5·10$^6$ to 1.5·10$^8$ cells.mL$^{-1}$. The same freezing, sampling and the subsequent OD measurement procedures described above were applied. Three replicates were performed for each cell density.

*Cell viability assay* – Cell viability assay was performed by plate counting. The frozen samples (*ca.* 300 µL) obtained after directional freezing were taken out from the -80 °C freezer and thawed in a water bath at 37 °C for 15 seconds. Subsequently, 120 µl of the samples were centrifuged at 6000 rpm and resuspended in 200 µl 0.9 wt.% NaCl solution. Samples were diluted to 1/10, 1/100 and 1/1000X. For plate counting, 100 µl of each of the diluted samples were pipetted onto the YPD agar surface (three replicates for each condition) and spreaded with a sterile plastic cell spreader. Agar plates were then put inside an incubator at 30 °C for 24 h and cell colonies were counted by visual inspection.

*SEM* – After directional freezing, samples were freeze-dried (ALPHA 2-4 LD, Bioblock Scientific, 0.1 mbar, at condenser temperature -80 °C) for 24 hours. The dried foams were cut perpendicular to the ice front growth direction (revealing the pores' cross section) and parallel to the ice growth direction (revealing the pores' alignment) and fixed to a metal SEM holder using carbon tape. Samples were sputtered with 10 nm gold layer and imaged on a Hitachi S-3400N scanning electron microscope at 10 kV beam acceleration voltage. Subsequent SEM image analysis were performed with FIJI software[45].

*Acknowledgments*

The authors show their appreciation to I. Génois for technical assistance on SEM and to C. Lorthioir and C. Boissière for fruitful discussions. KQ acknowledges funding from the China Scholarship Council, PhD grant n. 201606230232. This work was supported by French state funds managed by the National Research Agency (ANR) through the CellsInFoams project, grant n. ANR-17-CE08-0009 and the European Research Council under the European Community's Seventh Framework Programme (FP7/2007-2013) Grant Agreement n. 278004 (project FreeCo).

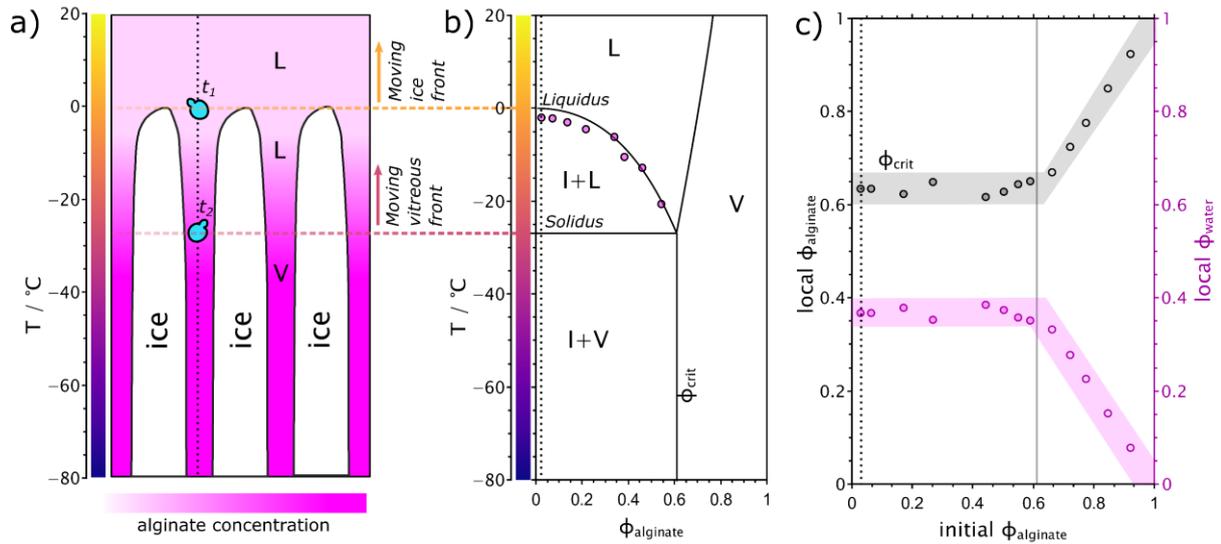

**Figure 1.** Phase segregation of alginate/water solution (*L*) during directional freezing enables controlled phase separation into ice (*I*) and a vitreous phase (*V*). **a)** Diagram of the freezing profile and phase separation during ice growth. At $t_1$, cells are brought into contact with the freezing front and enter the interstitial space defined by ice crystals. Between t₁ and t₂ cells experience an increasingly concentrated solution phase that co-exists with ice. When the composition and temperature of the interstitial space attain $\phi_{crit}$, defined as the intersection between the liquidus and $T_g$ lines, the cells become entrapped in a vitreous phase. **b)** Phase diagram of alginate/water solution. Magenta points describe the cryoscopic depression of alginate/water mixtures as determined by DSC at 10 °C.min⁻¹. Liquidus curve is defined by equation 2 and theoretical $T_g$ calculated according to Gordon Taylor model is defined by equation 3. **c)** The composition of segregated phase upon freezing according to initial alginate concentration. Non-freezing water contained in the segregated phase, in magenta, is determined by the difference between the total amount of water in the solution and the amount of water determined by the melting enthalpy of the frozen fraction. In black, alginate composition in the segregated phase according to the initial alginate volume fraction. © (2020) Qin et al. (10.6084/m9.figshare.12129117) CC BY 4.0 license https://creativecommons.org/licenses/by/4.0/.

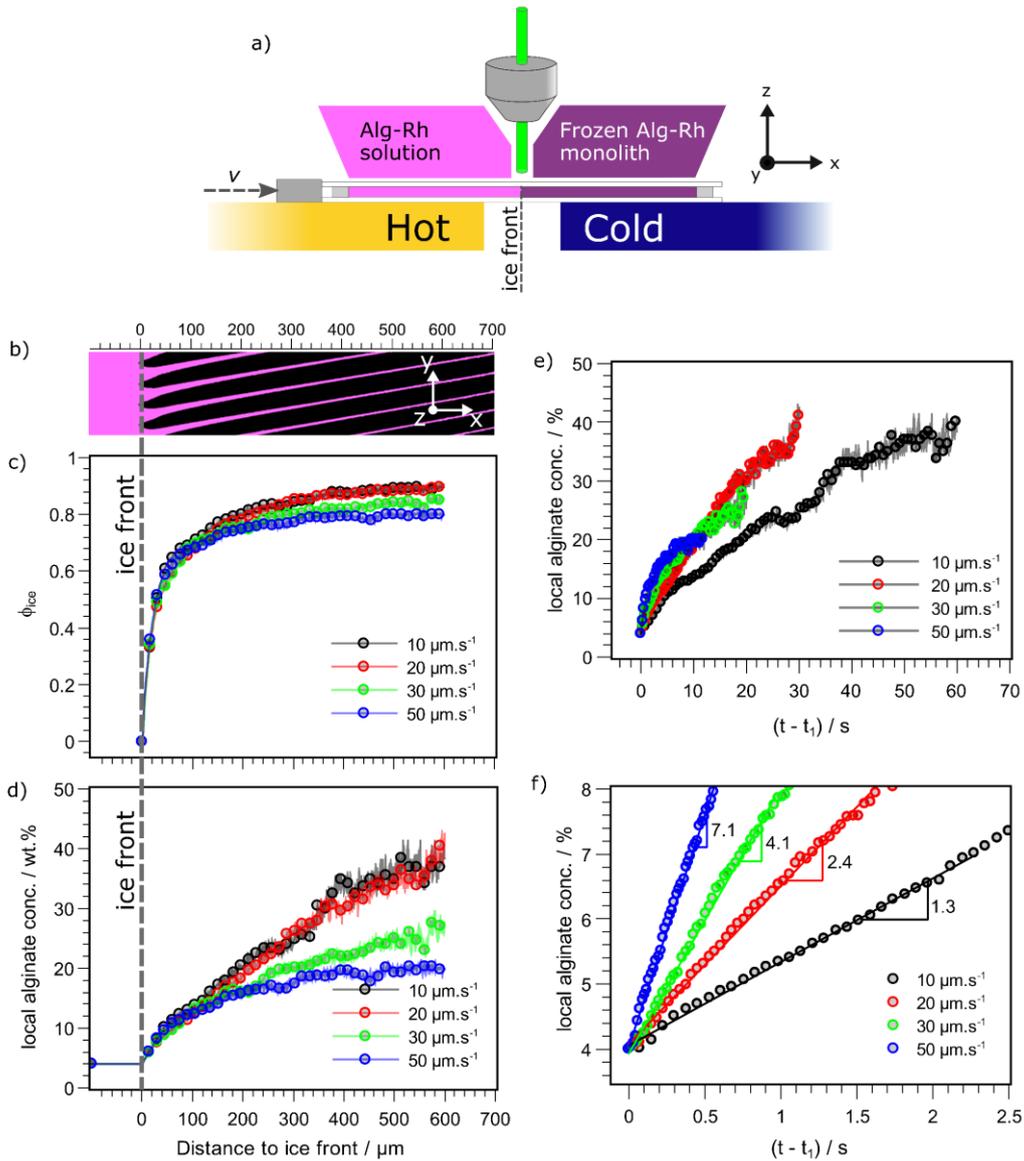

**Figure 2.** The composition of the interstitial space between ice crystals is determined by the linear freezing velocity. **a)** The freezing setup available under the confocal microscope enables to follow the freezing front at different linear velocities ranging from 10 to 50 µm.s⁻¹. **b)** Frame of the freezing front under the confocal microscope. In magenta, rhodamine B-grafted alginate; in black, ice. **c)** Ice volume fraction varies with the distance to the freezing front. Alginate/water segregated volume between ice crystals according to the ice front distance. Data is obtained from the integration of the volµme of ice crystals from the confocal microscopy obtained at 10 µm.s⁻¹, placed above the graph. **d)** Alginate concentration can be calculated as a function of the distance to the freezing front assuming a zero solubility in the ice fraction. e) Local alginate concentration evolution over time after interaction with the ice front ($t_1$). f) Initial moments of the alginate concentration evolution over time after interaction with the ice moving front. © (2020) Qin et al. (10.6084/m9.figshare.12129117) CC BY 4.0 license https://creativecommons.org/licenses/by/4.0/.

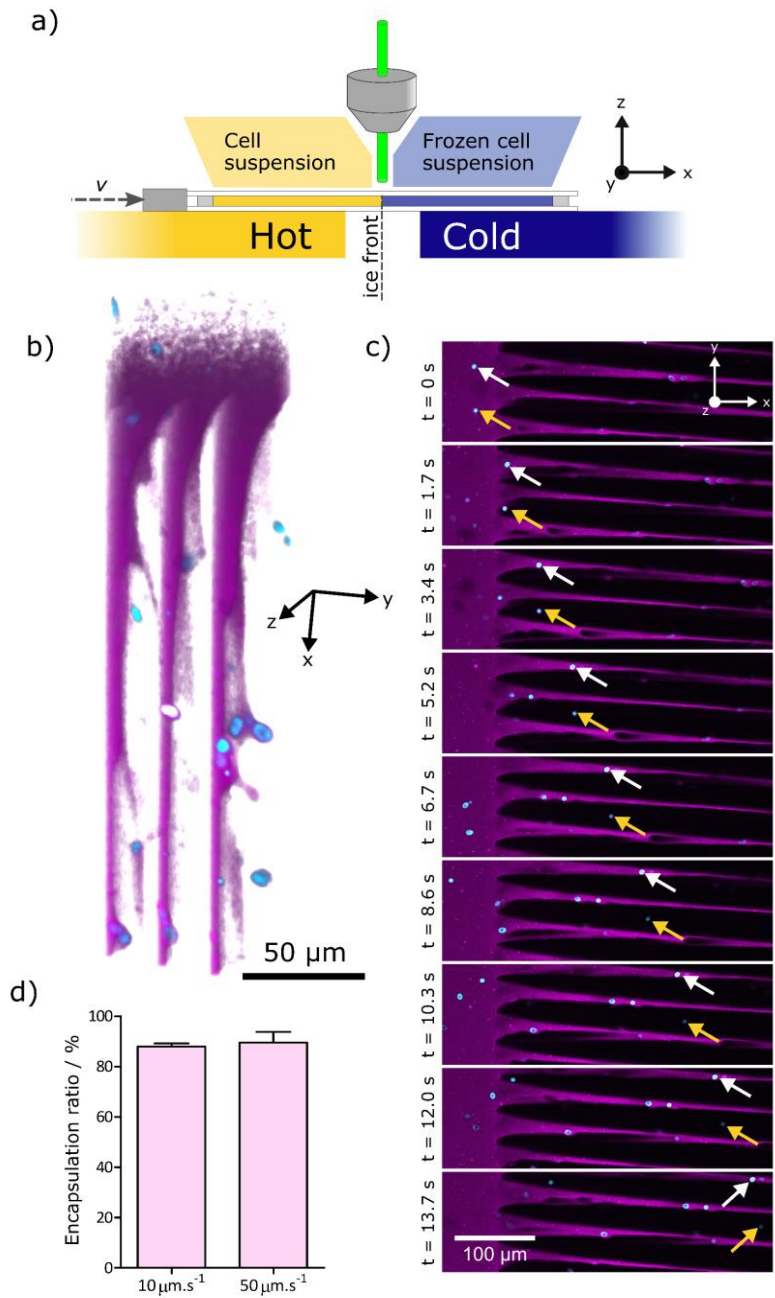

**Figure 3.** *In situ* directional freezing observation using confocal microscopy enables tracking the fate of individual *S. cerevisiae* cells at different ice front velocities. **a)** Directional freezing stage used to freeze cells under the confocal microscope. **b)** Three-dimensional rendered image of the freezing front region of a yeast cell suspension in alginate. Yeast cells (cyan) was stained with FUN1 dye, alginate was stained with Rhodamine B (magenta) and ice was unmarked (same color as background, white). **c)** Sequential 2D fluorescence imaging of the freezing front of yeast cells suspension frozen at 10 µm.s⁻¹. Fluorophore exclusion zones are occupied by ice (black). **d)** Statistic analysis of encapsulation efficiency for individual yeast cells at 10 and 50 µm.s⁻¹ freezing front velocity. © (2020) Qin et al. (10.6084/m9.figshare.12129117) CC BY 4.0 license https://creativecommons.org/licenses/by/4.0/.

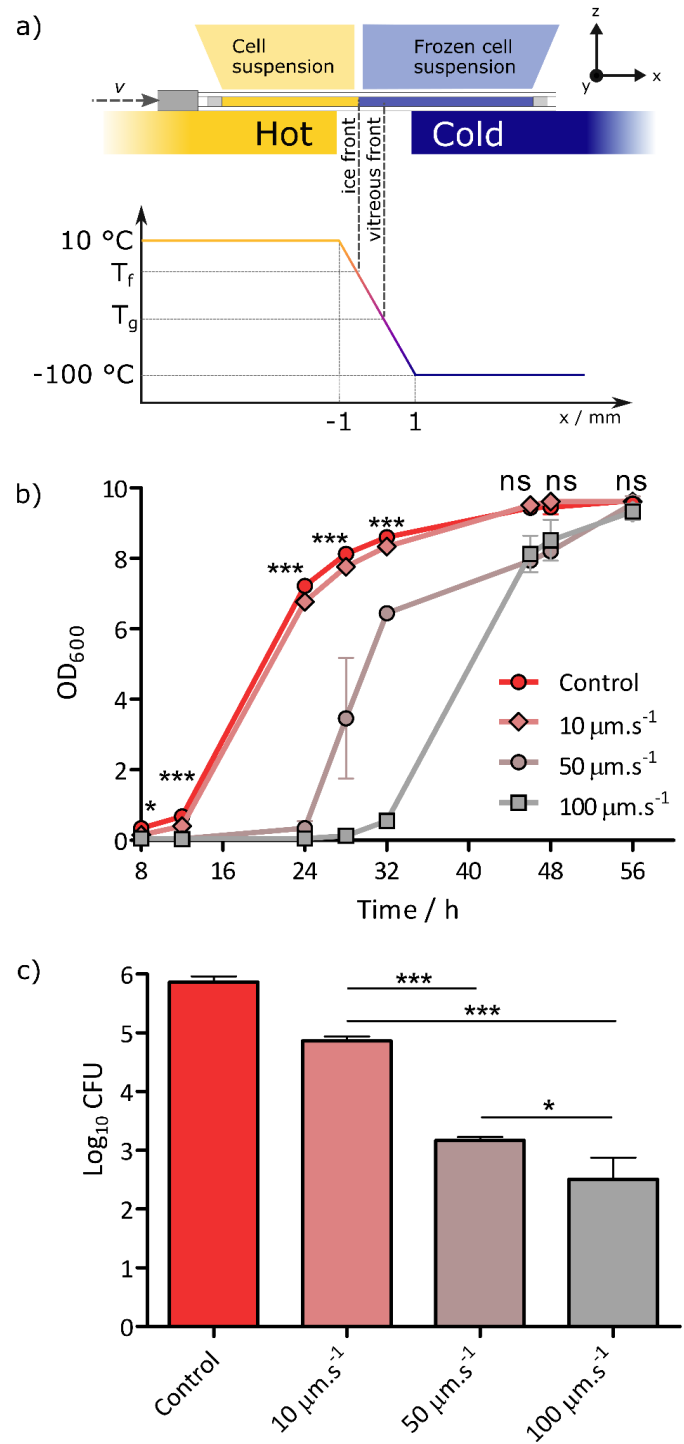

**Figure 4.** Ice front velocity determines yeast viability. **a)** Directional freezing setup used for freezing yeast cells suspended in alginate. **b)** Optical density of yeast suspension after directional freezing, thawing and re-culture in YPD suspension. **c)** Colony forming units of *S. cerevisiae* determined by plate counting performed immediately after thawing. © (2020) Qin et al. (10.6084/m9.figshare.12129117) CC BY 4.0 license https://creativecommons.org/licenses/by/4.0/.

# Unveiling cells' local environment during cryopreservation by correlative *in situ* spatial and thermal analyses

Kankan Qin[a], Corentin Eschenbrenner[a], Felix Ginot[b], Dmytro Dedovets[b], Thibaud Coradin[a], Sylvain Deville[b, c], Francisco M. Fernandes[a]*

[a] *Sorbonne Université, UMR 7574, Laboratoire de Chimie de la Matière Condensée de Paris, F-75005, Paris, France*

[b] *Laboratoire de Synthèse et Fonctionnalisation des Céramiques, UMR 3080 CNRS/Saint-Gobain CREE, Saint-Gobain Research Provence, Cavaillon, France*

[c] *now with: Université de Lyon, Université Claude Bernard Lyon 1, CNRS, Institut Lumière Matière, 69622 Villeurbanne, France*

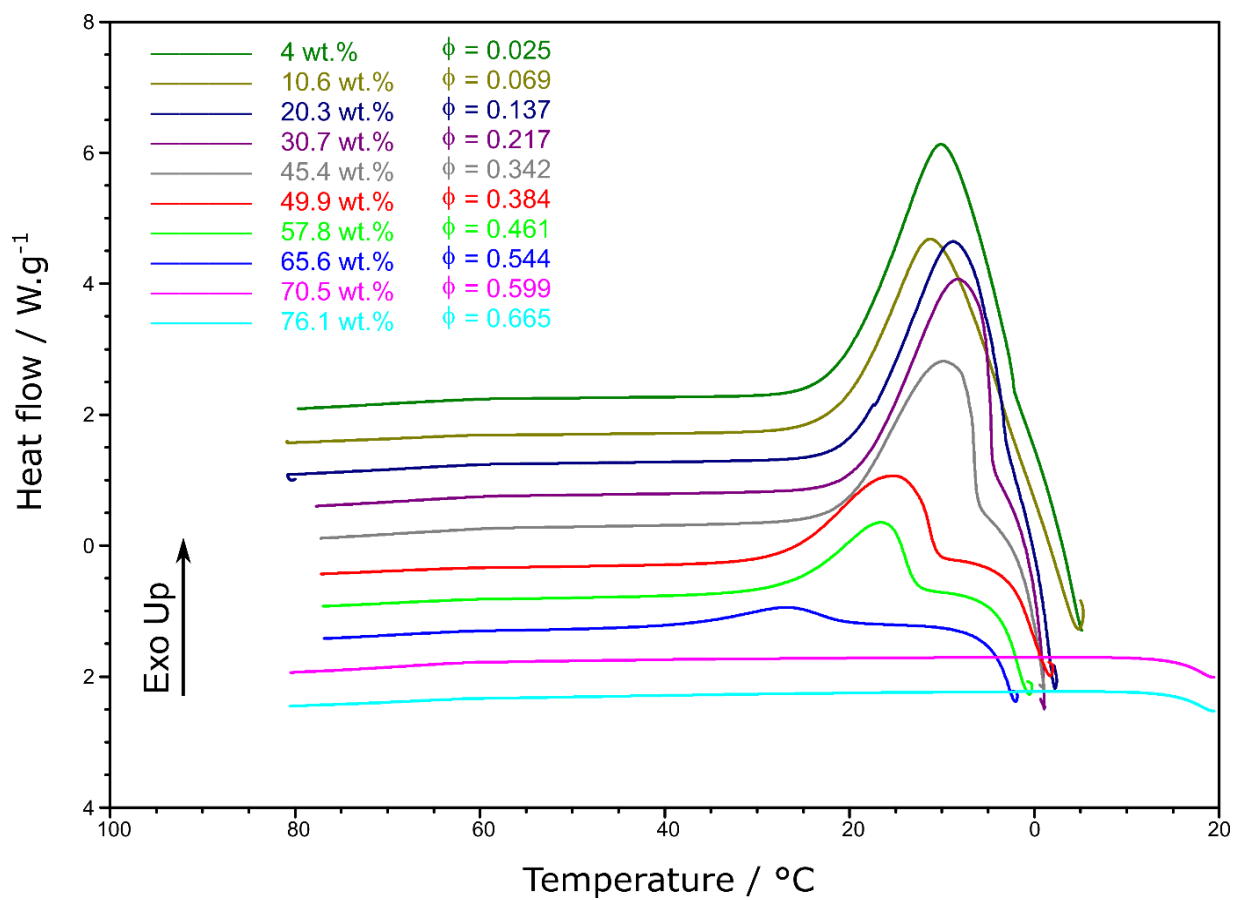

**Figure S1.** Cooling Differential Scanning Calorimetry scans obtained at 10 °C.min⁻¹ for alginate/water binary mixtures ranging from 4 to 76.1 wt.% in alginate.

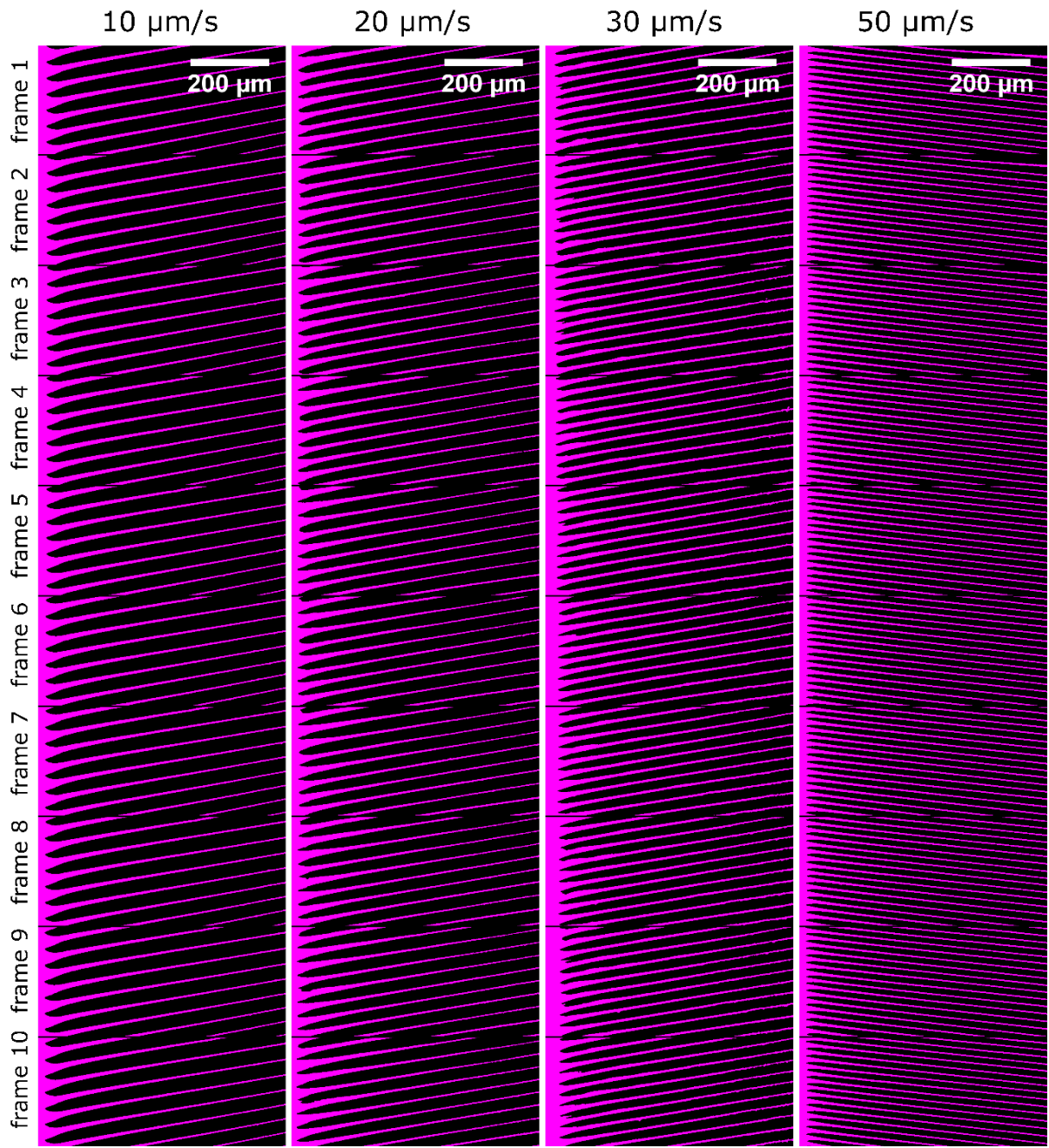

**Figure S2.** Examples of sequential frames used for the extraction of ice volume fraction at different ice front velocities

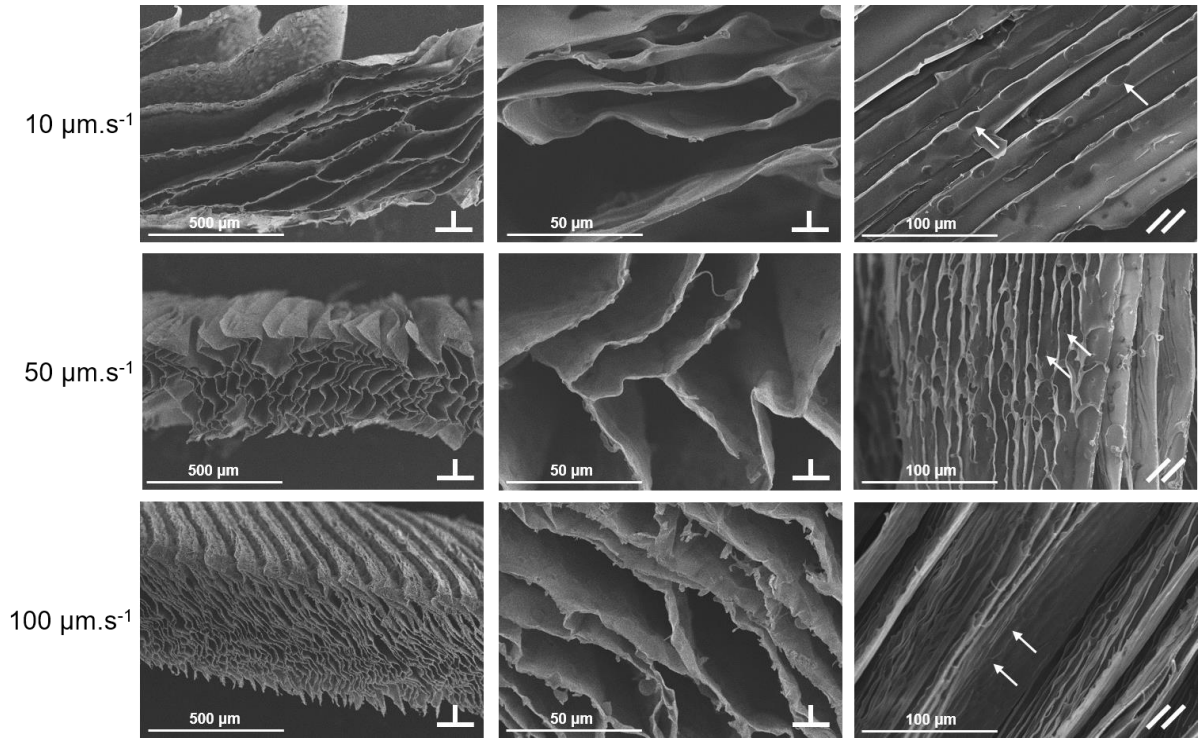

**Figure S3.** SEM micrographs of lyophilized alginate/cell foams prepared by directional freezing at different ice front velocities. Sample sections were cut perpendicular to the ice front growth direction. White arrows, indicate encapsulated yeast cells.

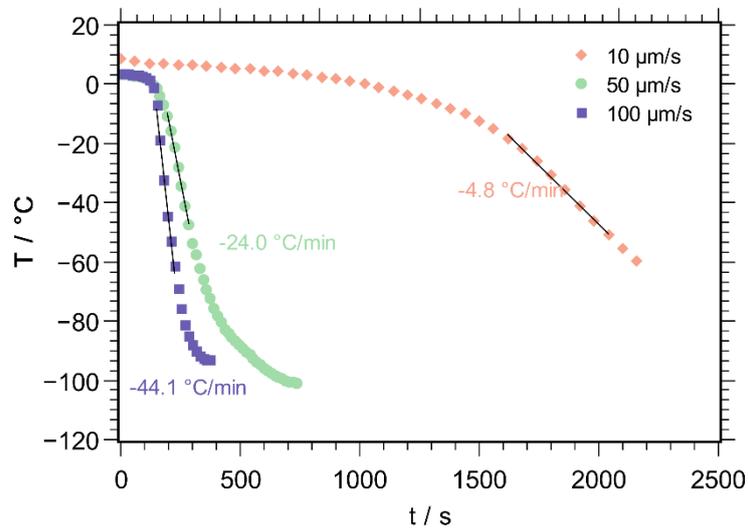

**Figure S4** Temperature variation at different ice front velocities in the home-made directional freezing setup.

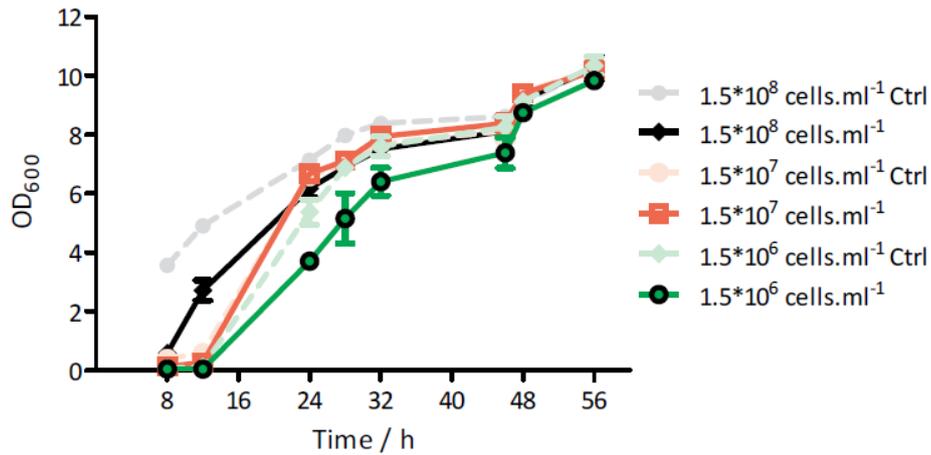

**Figure S5.** *S. cerevisiae* cell growth profiles with different initial cell densities (from $1.5 \cdot 10^6$ to $1.5 \cdot 10^8$ cells.mL$^{-1}$). Dot and light line, indicating control group without being frozen. Solid and dark line, indicating experimental group after subsequently freezing (10 µm.s$^{-1}$), thawing and re-culturing.

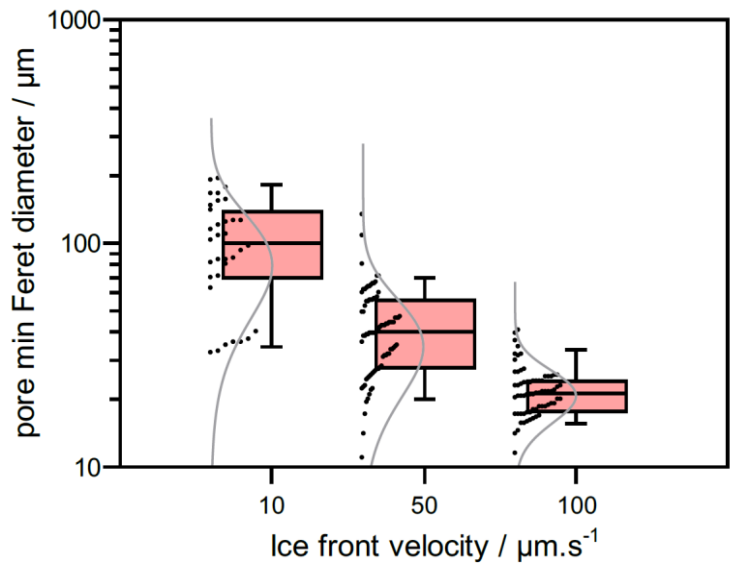

**Figure S6.** Minimum Feret diameter of pores performed on binarized images using FIJI software according to SEM images in Figure S3. Moustache boxplots depict the median (center), 1st and 3rd quartiles (bottom and top of boxes, respectively) and 10th and 90th percentile of the distribution (lower and superior moustache limits, respectively) for each ice front velocity. Black dots correspond to individual measurements (N>36). Fitting gamma function are juxtaposed with the boxplots and depicted to the left of each box.

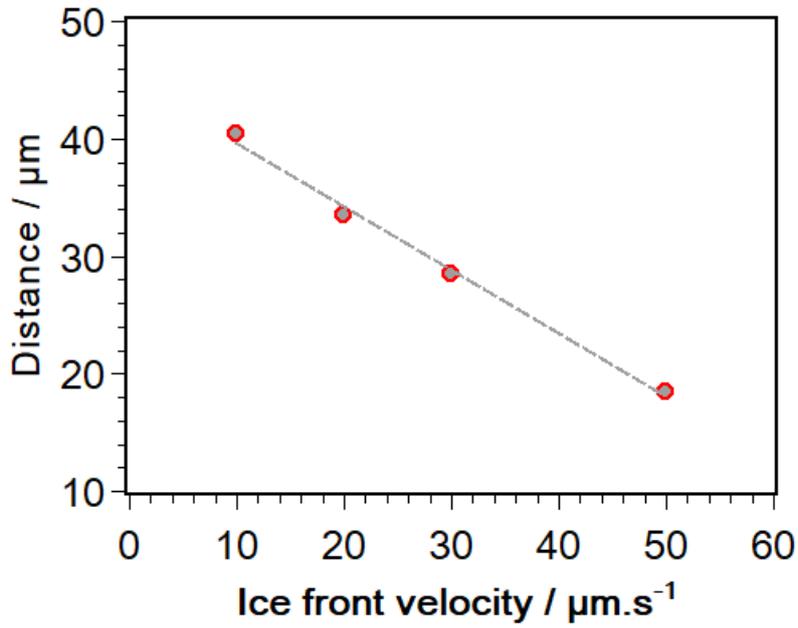

**Figure S7.** Periodic distances obtained by FFT analysis of confocal image stacks (n > 10) at different ice front velocities.

**Conversion between mass and volume fraction**

The conversion between mass and volume fraction assuming a zero volume change of mixing is given below,

$$\emptyset_p = \frac{\rho_w}{\rho_p\left(\frac{1}{x_w^{wt}-1}\right)+\rho_w} \qquad (S.eq.\ 1)$$

where $x_w^{wt}$ is the water mass fraction and $\rho_w$ and $\rho_p$ are the water and polymer density, respectively.

**Confocal image treatment for volume fraction analysis**

Image sequences of at least 10 consecutive frames were treated with a 2pixel median filter and rotated so that the ice front was horizontal. A 600*600 Gaussian blur filter was applied to each frame to extract a background intensity mask. Initial images were then divided by the obtained masks. A drift correction was applied using a movie stabilization macro by Nicholas Schneider (available from https://github.com/NMSchneider/fixTranslation-Macro-for-

[ImageJ/blob/master/NMS_fixTranslation_ver1.ijm](ImageJ/blob/master/NMS_fixTranslation_ver1.ijm)) to stabilize the ice moving front. Image sequences were cropped to remove stabilization artifacts at the image edges. All images extended for at least 600 µm below the ice front. Image sequences were subsequently binarized using a Huang threshold between 0 and 39, rotated by 90° and their profile integrated.